# Quantitative Analysis on the Disparity of Regional Economic Development in China and Its Evolution from 1952 to 2000

**Xu Jianhua**
East China Normal University
**Ai Nanshan**
Sichuan University
**Lu Yan**
East China Normal University
**Chen Yong**
Sichuan University
**Ling Yiying**
East China Normal University
**Yue Wenze**
East China Normal University

**Abstract.** Since late 1970s scholars have done much research on it, but conclusions from different scholars may differ in many ways. It is mainly due to different analytic approaches, perspectives, spatial units, statistical indicators and different periods for studies. On the basis of previous analyses and findings, we have done some further quantitative computation and empirical study, and revealed the inter-provincial disparity and regional disparity of economic development and their evolution trends from 1952—2000. The main conclusions are: (a) Regional disparity in economic development in China, including the inter-provincial disparity, inter-regional disparity and intra-regional disparity, has existed for years. (b) Gini coefficient and Theil coefficient have revealed a similar dynamic trend for comparative disparity in economic development between provinces in China. From 1952 to 1978, except for the "Great Leap Forward" period, comparative disparity basically assumes a upward trend and it assumed a slowly downward trend from 1979 to1990. Afterwards from 1991 to 2000 the disparity assumed a slowly upward trend again. In other words, the strategy of regional balanced development before the reform and opening up did not bring us a reduction in comparative disparity of regional economic development, nor did the lopsided development strategy implemented since then bring us an expansion of comparative disparity of regional economic development in China. (c) A comparison between Shanghai and Guizhou shows that absolute inter-provincial disparity has been quite big for years. The disparity of economic development between the two provinces expanded till 1978 and reduced after the reform and opening up. Since 1990 the disparity began to expand for the second times with a slight drop in 1998. (d) The R/S analysis result tell us that In the "Great Cultural Revolution" period, i.e. 1966-1978, the Hurst exponent H=0.504 0.5, indicates that in this period the evolution of comparative inter-provincial disparity of economic development showed a random characteristic, and In the other period, i.e. 1952-1965, 1979-1990 and 1991-2000, the Hurst exponent H>0.5 indicates that in this period the evolution of the comparative inter-provincial disparity of economic development in China has a long-enduring characteristic.

# RESEARCH BACKGROUND

Since late 1970s, more and more foreign scholars have started to study the disparity of regional development in China. N. R. Lardy[20/,21/] studied the output and income disparity between rural and urban areas, between agriculture and industry, and between inland and coastal regions before China's reform and opening up to the outside world and found that there was no exact fact showing the expansion of income disparity in different regions in China due to limit of data and information. Studies by C. Riskin[5/] and V. D. Lippit[24/] show that comparative disparity in income between provinces has drastically reduced. Compared with other developing countries, China had achieved remarkable progress, especially in social security. In contrast, J. Friedman[11/] and M. Selden[35/] et al argue that prior to China's reform and opening up disparity of regional development in China had been expanding. Apparently these scholars have a big difference in understanding disparity of regional development in China. Moreover, because of the low reliability of data they used, their conclusions lack credibility.

Since 1980s, along with release of substantial official data in China and improvement of research methodologies, some scholars have studied the evolution of regional development disparity in China since China's reform and opening up[19/,13/]. Among them P. Aguighier[1/] and D. Yang[45/], who was the first to study regional disparity since late 1970s, analyzed the strategic modes of regional development in China and its evolution. They argue that a bias policy in development enlarged development disparity between coastal and western region. K. Y. Tsui[38/,39/] analyzed average per capita national income (NI), and found that disparity of regional development in China hadn't much change during the period 1952-1970, but in the period 1970-1985 the disparity expanded.

T. P. Lyons[28/] was the first to use data released by China's State Statistical Bureau (SSB). He analyzed the change of disparity in average per capita net output value in each region of China from 1952 to 1987. He discovered comparative disparity had expanded in the periods of the "Great Leap Forward" (1958-1960) and the "Great Cultural Revolution" (1966-1976), but reduced in the period 1978-1987.

After 1990s, a series of new methods was adopted to further analyze the constitution and source of Chinese regional disparity. S. Rozelle[34] argued that regional disparity had greatly expanded in the coastal provinces from 1984 to 1989. When the Gini coefficient was decomposed he discovered development in rural industrialization was the main reason for expansion of regional disparity. When decomposing the Gini coefficient and the Theil coefficient, R. Kanbur and X. Zhang[17] found that from 1983 to 1995 the disparity between the rural and the urban was bigger than that between coastal and interior provinces. T. J. Kim and G. Knaap[18] studied regional disparity in agriculture, industry, construction industry and transportation by using Theil coefficient. The result indicated that from 1952 to 1985 disparity of costal provinces in agriculture, industry, construction industry and transportation contributed more than that of interior region in the same sectors to the overall disparity. Furthermore they argued this mode was obvious in late 1970s, but it was not promoted by China's strategy of economic development. Masahisa Fujita and D. Hu[32] decomposed the Theil coefficient with GDP and gross industrial output value, and came to the conclusion: disparity between coastal and interior provinces had been expanding. Although development disparity of coastal provinces was reducing, industrial development in coastal regions still developed fast. Moreover, he discussed the reasons for evolution of regional disparity in perspective of policies in regional development, economic globalization and liberalization. T. P. Lyons[29] focused his study in a smaller area. He analyzed the evolution of regional disparity in Fujian Province on a county level from 1978 to 1995. He discovered that the interior disparity of Fujian Province was expanding in terms of both absolute disparity and comparative disparity.

Later G. Long and M. K. Ng[26] also studied regional disparity in economic, social and cultural development in Jiangsu Province on a county level. They found the disparity had expanded since 1978. Besides, they analyzed the political, economic and social factors causing expansion of disparity. By using Solow's growth model, Chen and Fleisher[8] found that there is a conditional convergence for provinces in China for growth of per capita GDP in 1978-1993. They argued that regional disparity in China has experienced a reduction since implementation of the policy of reform and opening up.

Since 1990s many Chinese scholars have studied regional development disparity in China. Studies by some scholars show that the evolutional process of regional economic disparity in China is a "U-shaped" pattern. K. Yang[46] worked out the coefficient of weighed variation by per capita GDP at a provincial level. The conclusion obtained was: the evolution of provincial economic disparity in China was approximately in the shape of an inverted U-curve with an inflexion in the year of 1978. Prior to 1978 the disparity had seen an expansion and it began to decline after 1978. L. G. Ying[49] decomposed the Theil coefficient with the per capita GDP, and the results revealed a "U-shaped" pattern[44] in regional disparity from 1978 to 1994: before 1990 the regional disparity between coastal and interior provinces declined, after 1990 the disparity began to expand. By studying regional disparity based on income, D. Song[36] found national regional disparity was in the shape of an inverted " U " curve. Prior to 1990 there was a reduction for regional disparity, but after 1990 regional disparity gradually expanded. But not every one agrees upon the opinion that the evolutional process of regional economic disparity in China is a "U-shaped" pattern. W. Yang[47] calculated the Gini coefficient with per capita GNP, and analyzed evolution of income disparity between coastal, middle and western regions of China in the 1980s. The conclusion was: China's biased strategy to give priority to coastal regions for development had led neither to expansion of income disparity all over china, nor to expansion of income disparity between coastal, middle and western regions in China. On the contrary there was an overall decline of income disparity in China. H. Wei[40,41,42] analyzed disparity evolution of the three supra-provincial regions in the period 1978-1992. The result indicated that all the three supra-provincial regions had enhanced in their economic strengthen. The middle and western region however obviously lagged behind in terms of development pace. The disparity between coastal and middle-western regions was still in expansion. Wei and Liu[43] also forecast the economic development trend of the three supra-provincial regions: from 1993 to 2010, economic growth would remain unbalanced. Absolute disparity between coastal and middle- western regions wouldn't reduce, and comparative disparity might expand in the near future. D. Lu and F. Xue, et al.[27] and A. Hu and P. Zou[14] argued that overall disparity of regional development in China had expanded before 1978 and then began to reduce until 1990s when disparity once again saw an increase. G. Yuan[50] believed a remarkable trend in regional development of China since the reform and opening up was the enlarging disparity between the three supra-provincial regions, which is in conformity with evolution of overall economy in China. Y. Lin and F. Cai et al.[23] have studied evolution of regional disparity during China's economic transition period (1978-1995) by using per capita GDP and per capita income. It was found that disparity between the three supra-provincial regions have a greater effect on China's overall disparity than the intra-regional disparity. F. Cai and Y. Du[3,4] broke down national overall disparity into the interregional disparity and intra-regional disparity and found that in the period 1978-1999 the intra-regional disparity of coastal region had a major contribution to the overall disparity, but in a declining trend; the intra-regional disparity of the middle region contributed a little to the overall disparity, and in a declining trend as well; the intra-regional disparity of the western region contributed very little to the overall disparity, and also in a declining trend; the interregional disparity between the three supra-provincial regions contributed substantially to the national overall disparity, and in a marked increasing trend. They argued that there was a conditional similarity in economic growth in different regions in China. Using nationwide county level data, X. Li and J. Qiao[22/] analyzed, for the first time at county level, the spatial evolution of regional economic disparity in China in the 1990s. The results demonstrated that economic disparity between counties had declined, but the disparity between coastal and inland regions had widened; The counties with faster economic growth than the national average level were chiefly distributed in three growth belts, namely, the coastal belt (along China's coastal line), Beijing-Guangzhou railway belt (along the railway from Beijing to Guangzhou) and the Yangtze River belt (along the Yangtze River from Chongqing to Shanghai). The less developed counties were however mainly located in the western part of China. Q. Liu[25/] argued convergence of regional economic growth in China appeared in different time and at different regions, and disparity in output between regions was positively correlated to overall economic instability in China.

From above analysis, as we know, domestic and foreign scholars have already done much research on regional disparity and its evolution in China, but there is a big difference in conclusions. What is the reason for this? We think it is mainly due to different analytic approaches, perspectives, spatial units, statistical indicators and different periods for studies. On the basis of previous analyses and findings, we have done some further quantitative computation and empirical study, and revealed the inter-provincial disparity and regional disparity of economic development and their evolution trends from 1952—2000.

## METHODOLOGY AND DATA

### The division of the spatial unit

There is usually a spatial criterion for studies of regional disparity. What spatial criterion should we choose for researches? It depends on study purposes and specific objectives. The purpose of this paper is to reveal provincial disparity, regional disparity in economic development from 1952—2000 and their evolution. Provincial administrative unit is a political and economic region with an integrated function, and each with a complete system of statistical data, which is readily available. Thus, we choose the provincial district (provinces, municipalities and autonomous regions) as basic spatial unit for our analysis, and also choose the three supra-provincial regions: coastal, middle and western as more overall spatial units. The coastal provinces are Beijing, Tianjin, Hebei, Liaoning, Shanghai, Jiangsu, Zhejiang, Fujian, Shandong, Guangdong, Guangxi and Hainan; The middle provinces Shanxi, Inner Mongolian, Jilin, Heilongjiang, Anhui, Jiangxi, Henan, Hubei and Hunan; The western provinces are Yunnan, Guizhou, Sichuan, Chongqing, Tibet, Shaanxi, Gansu, Qinghai, Ningxia and Xinjiang (map 1).

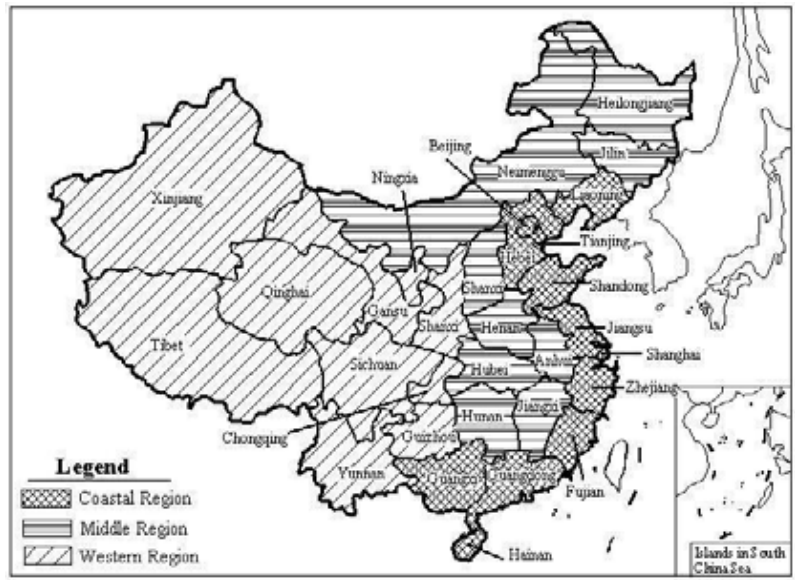

Figure1. The division of the spatial unit of China

## Selection of statistical indicators and sample data

As for study of dynamic evolution of regional disparity in China, per capita GDP of each province may be appropriate. Since per capita GDP is the best approximation and can well reflect the overall development level and people's well being, it is widely used. Moreover, the time series data in per capita GDP in each province is complete and can be used for temporal and spatial comparison. Therefore, we choose 31 provinces (municipalities, autonomous region) in China as spatial samples, the period 1952-2000 as temporal samples.

The primary data published by National Bureau of Statistics of China, are mainly from: (1) Comprehensive Statistical Data and Materials on 50 Years of New China. Beijing: China Statistics Press, 1999. (2) China Statistical Yearbook 2001. Beijing: China Statistics Press, 2001. (3) Urban Statistical Yearbook of China 2001. Beijing: China Statistics Press, 2001. (4) Historical Data for China Gross Domestic Product (1952-1995). Dalian: Northeast China University of Finance and Economics Press, 1997. In a general way, the data are creditable and authoritative theoretically. However, several years' data such as those during the "Great Leap Forward" period are distorted because of none-economic factors, which is proved by Gini and Theil coefficient in section 3.1 in this paper. Consequently, these data are not so credible and the result calculated by using them is not so precise. However, as we study regional economic disparity in China in as long period, the primary data can only from National Bureau of Statistics of China.

## Data processing method for eliminating influence of prices

If the price of products and services does not change at all, we can ignore the influence from price change. China is however a large country with a big spatial difference. During its past 50 years of economic development, China has undergone several different stages when the comprehensive price index and inflation rate was constantly changing in different periods and at different places. Therefore, if we discuss regional development disparity using present price, we may have an erroneous conclusion. In order to accurately reflect the disparity and its evolution, we must consider the influence of the price factor. So we convert all GDP data of each region into present value by using price index (1) as follows[14/].

$$X_i(t) = X_i(t_0) \times X_i^E(t) \qquad 1$$

in which $X_i(t)$ is the real GDP data of the $i$ th region at the $t$ th year, $X_i(t_0)$ the real GDP data of the $i$ region at the first period (the $t_0$ th year), $X_i^E(t)$ is the GDP growth exponent of the $i$ region at the $t$ th year.

In this article 1978 is the base year, and the GDP data of each year was the real data converted in term of comparable price.

## Quantitative analysis methods

**Quantitative methods to describe regional disparity.** There are so many quantitative exponents used to describe regional disparity[2/,6/,9/,10/,12/,33/,44/,48/], such as extreme deviation, standard deviation, coefficient of variation, Engel coefficient, location entropy and so on. By comparing all these methods, we choose Gini coefficient and Theil coefficient as quantitative indicators for analysis of disparity in regional economic development in China. Moreover these two indicators are commonly used in present study of regional economics worldwide.

**Gini coefficient.** In the actual applications, according to characteristic of statistical data, there are several kinds of methods for computerizing Gini coefficient. Curve fitting, as one of the simplest ones, is done as follows:

Assuming that $X$ is the accumulated percentage of population of the regions with a per capita GDP less than a certain level in total population in China. Y is the accumulated percentage of GDP of the regions with a per capita GDP less than a certain level in total GDP in China. The functional relation between $X$ and $Y$ ( Lorentz equation) is

$$Y = f(X) \qquad 0 \leq X \leq 1 \qquad 2$$

Then, the Gini coefficient (G) is defined as

$$G = 1 - 2\int_0^1 f(X)dX \qquad 3$$

If the equation is fitted with the power function $Y = X^\beta$, it follows that

$$G = \frac{\beta - 1}{1 + \beta} \qquad 4$$

where $\beta$ can be fitted with the least square method (LSM),

$$\beta = \frac{\sum_{i=1}^{k}(\ln X_i \times \ln Y_i)}{\sum_{i=1}^{k}(\ln X_i)^2} \qquad 5$$

in which $Y_i$ is the accumulated percentage of the $i$th region's per capita GDP, $X_i$ the respected accumulated percentage of population.

If the Gini coefficient is bigger, disparity in economic development between regions will be bigger. Otherwise, disparity will be smaller.

**heil coefficient.** Theil coefficient is also called Theil entropy[37/], which was proposed by Theil H. in 1967. Theil coefficient is defined in the following way:

$$T = \sum_{i=1}^{N} Y_i \log \frac{Y_i}{P_i} \qquad 6$$

where $N$ is the number of areas, $Y_i$ is the share of $i$th region in total GDP in the whole country, $P_i$ is the share of $i$th region in total population in the whole country.

If Theil coefficient is bigger, disparity in economic development between various areas will be bigger. Otherwise, disparity will become smaller.

**Rescaled range analysis on evolution of regional disparity**

All statistical methods assume that all data of time series be independent (i.e. fit for Gauss distribution), hence the series is stochastic. When H. E. Hurst[15/,16/], a British physicist, analyzed water level of the Nile River, he found that such time series like river water level was not fit for Gauss distribution, showing a characteristic of discontinuity and durability. Based on the empirical findings of H. E. Hurst, B. B. Mandelbrot made a breakthrough on fundamental theories of traditional statistical methods. He divided time series into two categories: discontinue and durable, of which the former is called Noah effects and the later called Joseph effects, both originating from Gensis 6 of Old Testament as" ……were all the fountain of the great deep broken up, and the window of heaven were opened. And the rain was upon the earth forty days and forty nights." The torrential rain lasting for 40 days and nights presents a characteristic of unevenness in rainfall and discontinuity in time. The story of Joseph in Gensis 41 reads like " ……there came seven years of great plenty throughout the land of Egypt. And there shall arise after them seven years of famine" showing the periodical occurrence of humidity and aridity with a characteristic of durability. The idea was proposed in his speech named " New forms of chance in the sciences" and recorded in references of B. B. Mandelbrot and J. R. Wallis[30/] and B. B. Mandelbrot[31/]. Under the effect of Noah effects and Joseph effects, time series no longer present a random Brownian movement unrelated to the past, but show a characteristic of long-term correlation, which was called fBM (fractional Brownian Movement) by B. B. Mandelbrot and could be studied by R/S analysis. The rescaled range (R/S) analysis, as a non-linear method for forecast, has been widely used in many researches[7].

The principle of R/S analysis is as following: Considering the time series $\{\xi(t)\}$ (t =1, 2,… ) of Theil coefficient (or Gini coefficient) variation, for any positive integer $\tau \geq 1$, the mean value series is defined as

$$\langle \xi \rangle_\tau = \frac{1}{\tau} \sum_{t=1}^{\tau} \xi(t) \qquad \tau = 1, 2, \ldots$$

The accumulative deviation is

$$X(t, \tau) = \sum_{u=1}^{t} (\xi(u) - \langle \xi \rangle_\tau) \qquad 1 \leq t \leq \tau$$

The extreme deviation is

$$R(\tau) = \max_{1 \leq t \leq \tau} X(t, \tau) - \min_{1 \leq t \leq \tau} X(t, \tau) \qquad \tau = 1, 2, \ldots$$

The standard deviation is

$$S(\tau) = \left[ \frac{1}{\tau} \sum_{t=1}^{\tau} (\xi(t) - \langle \xi \rangle_\tau)^2 \right]^{\frac{1}{2}} \qquad \tau = 1, 2, \ldots$$

When analyzing the statistic rule of $R(\tau)/S(\tau) \triangleq R/S$, Hurst discovered a relational expression

$$R/S \propto \left(\frac{\tau}{2}\right)^H \qquad (7)$$

It shows there is Hurst phenomenon in time series, and H is called the Hurst exponent. According to ($\tau, R/S$), H can be obtained by least square method (LSM) in log-log grid. Hurst et al. once proved that if $\{\xi(t)\}$ is independently random series with limited variance, the expression H=0.5, H (0<H<1) is dependent of an incidence function $C(t)$

$$C(t) = 2^{2H-1} - 1 \qquad (8)$$

When H>0.5, $C(t)$>0, it means that the future trend of time series will be consistent with the past. In other words, if the past disparity of regional economic development has been enlarged, the disparity in the future will also be enlarged. The process of regional economic development will assume a divergent trend; When H<0.5, $C(t)$<0, it means the future trend of time series will be opposite from the past. In other words, if the past disparity of regional economic development has an expansive trend, the disparity of the future will assume the contractive trend. The process of the regional economic development will assume a convergent trend; When H=0.5, $C(t)$=0, it means time series is completely independent. There is no correlation or only short-range correlation in time series and we cannot conclude whether the disparity of regional economic development will expand or contract. The process of regional economic development will neither converge nor diverge.

# THE EVOLUTIONARY PROCESS OF REGIONAL ECONOMIC DEVELOPMENT DISPARITY IN CHINA FROM 1952 TO 2000

## The relative disparity shown by Gini and Theil coefficient

Gini coefficient and Theil coefficient calculated from per capita GDP converted in the period 1952-2000 can reveal evolution of relative inter-provincial disparity since 1949 (Fig.1).

**Figure 2.  The Gini and Theil coefficient by comparative price from 1952 to 2000**

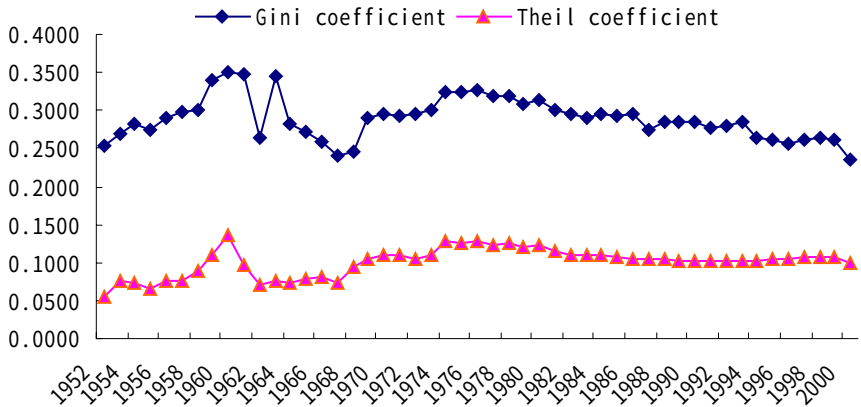

With reference to Fig.1, we can see that Gini coefficient and Theil coefficient revealed the same trend of the evolution of comparative inter-provincial disparity. From 1952 to 1978, except for several unusual data in the "Great Leap Forward" period, the disparity assumes the upward trend on the whole; From 1979 to 1990 the disparity assume the slowly downward trend. But from 1991 to 2000 the disparity assumes the slowly upward trend again. It is evident that the magnitudes of changes in disparities are much bigger during the earlier years than in the 1978-2000 period. In other words, while the strategy to balance regional development before the reform and opening up has not succeeded in reducing comparative disparity in China's regional economic development, the lopsided development strategy after 1978 also has not enlarged disparity. This conclusion is very interesting and exciting, and it gives us a new research task to explain the reasons for the bizarre phenomenon.

## A comparison between Shanghai and Guizhou: the provincial disparity between maximum and minimum

Gini coefficient, Theil coefficient and its decomposition have reflected comparative inter-provincial disparity, comparative inter-regional and intra-regional disparity in three supra-provincial regions, but they covered up the absolute inter-provincial disparity. Therefore, it is necessary to choose two provinces for comparison so as to reveal the evolution of the absolute inter-provincial disparity. In terms of this principle, we choose shanghai with highest economic development and Guizhou with the lowest economic development for comparison.

The change of ratios of per capita GDP in Shanghai to that in Guizhou Province not only reflects disparity between Shanghai and Guizhou, but also reveals, at some extent, absolute interregional disparity. It is clear in Fig.2 that evolution of absolute disparity between Shanghai and Guizhou has experienced three stages. The disparity of economic development between the two provinces expanded till 1978 and reduced after the reform and opening up. Since 1990 the disparity began to expand for the second times with a slight drop in 1998. Before 1978 the disparity between Shanghai and Guizhou expanded at a comparatively larger extent with an increase of disparity ratio from 12.355 (1952) to 28.076 (1978). After 1978 the disparity between two provinces reduced, with a disparity ratio of 24.026 in 1990. In 1990 the disparity started to expand again. In 1998 it began to reduce.

**Figure 3.** The ratios of per capita GDP in Shanghai Municipality to those in Guizhou Province from 1952 to 2000

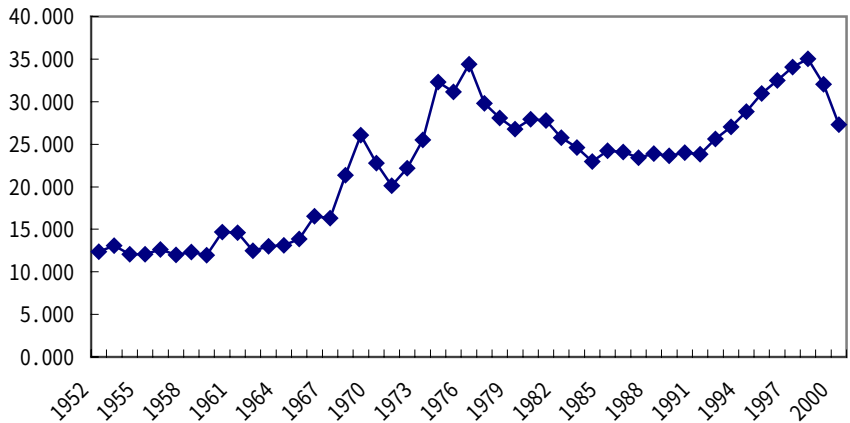

Compared with national average per capita GDP in the period 1952-2000(Tab.1), the ratio of Shanghai to China as a whole was all above 4 while the ratio of Guizhou to the whole China was less than 1, and even under 1/2. A comparison of development between Shanghai and China as a whole is approximately in conformity with a comparison between Shanghai and Guizhou in the following way: the disparity had increased prior to 1978 and it slightly reduced from 1978 to 1990. After 1990 it increased again until 1998 when there was a slight reduction. Before 1976 the disparity between Guizhou and China as a whole had increased and it slightly reduced from 1978 to 1990. Since 1990 the disparity has enlarged again.

**Table 1.** The ratios of per capita GDP of Shanghai Municipality and Guizhou Province to the **national average per capita GDP from 1952 to 2000**

| Year | 1952 | 1953 | 1954 | 1955 | 1956 | 1957 | 1958 | 1959 | 1960 | 1961 |
|---|---|---|---|---|---|---|---|---|---|---|
| Shanghai / the national average | 4.306 | 4.625 | 4.345 | 4.368 | 4.757 | 4.522 | 4.829 | 4.353 | 5.140 | 4.993 |
| Guizhou / the national average | 0.349 | 0.354 | 0.361 | 0.362 | 0.377 | 0.377 | 0.391 | 0.365 | 0.350 | 0.342 |
| Year | 1962 | 1963 | 1964 | 1965 | 1966 | 1967 | 1968 | 1969 | 1970 | 1971 |
| Shanghai / the national average | 4.119 | 4.356 | 4.468 | 4.740 | 4.933 | 4.976 | 5.778 | 5.815 | 5.634 | 5.627 |
| Guizhou / the national average | 0.331 | 0.335 | 0.341 | 0.342 | 0.298 | 0.305 | 0.270 | 0.223 | 0.247 | 0.279 |
| Year | 1972 | 1973 | 1974 | 1975 | 1976 | 1977 | 1978 | 1979 | 1980 | 1981 |
| Shanghai / the national average | 5.782 | 5.979 | 6.332 | 6.175 | 6.294 | 6.408 | 6.536 | 6.452 | 6.417 | 6.503 |
| Guizhou / the national average | 0.261 | 0.234 | 0.196 | 0.198 | 0.183 | 0.215 | 0.233 | 0.241 | 0.230 | 0.234 |
| Year | 1982 | 1983 | 1984 | 1985 | 1986 | 1987 | 1988 | 1989 | 1990 | 1991 |
| Shanghai / the national average | 6.429 | 6.264 | 6.072 | 6.129 | 5.991 | 5.878 | 5.841 | 5.812 | 5.814 | 5.780 |
| Guizhou / the national average | 0.249 | 0.254 | 0.264 | 0.253 | 0.249 | 0.251 | 0.244 | 0.246 | 0.242 | 0.243 |
| Year | 1992 | 1993 | 1994 | 1995 | 1996 | 1997 | 1998 | 1999 | 2000 | |
| Shanghai / the national average | 5.867 | 5.910 | 5.975 | 6.087 | 6.187 | 6.325 | 6.384 | 5.949 | 5.581 | |
| Guizhou / the national average | 0.229 | 0.218 | 0.207 | 0.197 | 0.190 | 0.186 | 0.182 | 0.186 | 0.204 | |

# R/S ANALYSIS RESULTS OF THEIL COEFFICIENT SEQUENCE: THE DYNAMIC OF POVINCIAL DISPARITY

While scholars both in China and abroad study disparity in regional economic development in China, they have been concerned a common issue, i. e., whether economic development in different regions of China will converge, or whether the income level in each region will have a convergence? Based on the Solow growth model, J. Chen and B. M. Fleisher[8/] studied China's regional disparity by using per capita GDP, and discovered that regional economic disparity from 1978 to 1993 in China showed a conditional convergence, i. e. it depended on the share of physical capital, employment growth, investment in human capital, foreign direct investment and location. On the contrary, S. Yao and Z. Zhang[48/] analyzed convergence of China's regional economy from 1952 to 1997 with per capita GDP. The result indicates regional disparity of economic development in China would expand. As for these conclusions, we have our own ideas. The following is our R/S analysis on this issue.

Taking Theil coefficients from 1952 to 2000 obtained in previous paragraphs as time series $\xi(t)$, we have calculated the Hurst exponent H according to temporal characteristics of economic development in China. The result is in Tab.2.

**Table 2. The Hurst exponent of the Theil Coefficient series from 1952 to 2000**

| Periods | 1952—1965 | 1966—1978 | 1979—1990 | 1991—2000 | 1952—2000 |
|---|---|---|---|---|---|
| Hurst Exponent | 0.670 | 0.504 | 0.722 | 0.730 | 0.545 |

With reference to Tab.2, We may reach the following conclusions:

(1) In the period 1952-1965, the Hurst exponent H=0.670>0.5 indicates that in this period evolution of comparative inter-provincial disparity of economic development showed a long-enduring characteristic. In the period 1966-1978, the Theil coefficient assumes an increasing trend, which has confirmed this conclusion.

(2) In the period 1966-1978, the Hurst exponent H=0.504 0.5, indicates that in this period the evolution of comparative inter-provincial disparity of economic development showed a random characteristic, because it is in the period of the "Great Cultural Revolution" so conclusion was verified.

(3) In the period 1979-1990, the Hurst exponent H=0.772>0.5 indicates that in this period the evolution of the comparative inter-provincial disparity of economic development between provinces in China has a long-enduring characteristic, in the period 1991-2000 Hurst exponent H=0.730>0.5, which has also verified the conclusion.

## CONCLUSIONS

From above modelling results, we elicit the following conclusions.

(1) Regional disparity in economic development in China, including the inter-provincial disparity, inter-regional disparity and intra-regional disparity, has existed for years.

(2) Gini coefficient and Theil coefficient have revealed a similar dynamic trend for comparative disparity in economic development between provinces in China. From 1952 to 1978, except for the "Great Leap Forward" period, comparative disparity basically assumes an upward trend and it assumed a slowly downward trend from 1979 to1990. Afterwards from 1991 to 2000 the disparity assumed a slowly upward trend again. In other words, the strategy of regional balanced development before the reform and opening up did not bring us a reduction in comparative disparity of regional economic development, nor did the lopsided development strategy implemented since then bring us an expansion of comparative disparity of regional economic development in China. This conclusion is very interesting and exciting, and it gives us a new research task to explain the reasons for the bizarre phenomenon.

(3) A comparison between Shanghai and Guizhou shows that absolute inter-provincial disparity has been quite big for years. The disparity of economic development between the two provinces expanded till 1978 and reduced after the reform and opening up. Since 1990 the disparity began to expand for the second times with a slight drop in 1998.

(4) The R/S analysis result tell us that In the "Great Culture Revolution" period, i.e. 1966-1978, the Hurst exponent H=0.504 0.5, indicates that in this period the evolution of comparative inter-provincial disparity of economic development showed a random characteristic, and In the other period, i.e. 1952-1965, 1979-1990 and 1991-2000, the Hurst exponent H>0.5 indicates that in this period the evolution of the comparative inter-provincial disparity of economic development in China has a long-enduring characteristic.

# NOTES

1/ Aguignier, P. 1988. Regional Disparity Since 1978, in Feuchtwang,Sephen, Hussain, Athar and Pairault, Thierry, (Ed.), *Transforming China's Economy in the Eighties: the Urban Sector*, 2, 93-106. London: Zed Books Ltd.
2/ Borts, G. H. Stein, J. L. 1964. *Economic Growth in a Free Market*. New York: Columbia University Press.
3/ Cai, F., Du Y. 2000. Convergence and divergence of regional economic growth in China. *Economic Research Journal* (in Chinese). 10: 30-37.
4/ ----------------. 2001. Regional gap, convergence and developing china's western region, in China. *China Industrial Economy* (in Chinese), 2: 48-54.
5/ C. Riskin. 1978. *China's Political Economy: The Quest for Development since 1949*. Oxford: Oxford University Press.
6/ Carlberg, M. 1981. A neoclassical model of interregional economic growth. *Regional Science and Urban Economics*. 11: 191-203.
7/ Chen, R. Wang, F. Li, H. Ai, N. 1992. The R/S analysis of China's population development. *Population Science of China* (in Chinese), 4: 27-32.
8/ Chen, J. & Fleshier, B. M. 1996. Regional income inequality and economic growth in China. *Journal of Comparative Economics*. 22 (2): 141-164.
9/ Dowrick, S., Nguyen, D. 1989. OECD comparative economic growth 1950-1985: catch-up and convergence. *American Economic Review*.
10/ Dunford, M. 1993. Regional disparities in the European community: evidence from the REGIO databank. *Regional Studies*. 27(8): 727-743.
11/ Friedman, E. 1987. Maoism and the liberation of the poor. *World Politics*. 39(3): 408-428.
12/ Friedman, J. 1963. Regional economic policy for developing areas. *Papers of the Regional Science Association*, 11.
13/ Jian, T., Sachs, J. D. Warner, A. M. 1996. Trends in regional inequality in China. *National Bureau of Economic Researsh Working Paper 5412*. Cambridge, MA.
14/ Hu, A. Zou, P. 2000. *Society and development: a study of China's regional social development disparity* (in Chinese). Hangzhou: Zhejiang People's Publishing House.
15/ Hurst, H. E. 1951. Long-term storage capacity of reservoirs. *Transactions of American Society of Civil Engineers* 116: 770-799.
16/ --------------- 1955. Methods of using long-term storage in reservoirs. *Proc. Of the Institution of Civil Engineers*, Part I: 519-577.
17/ Kanbur, R. Zhang, X. 1999. Which regional disparity? the evolution of rural-urban and inland-coastal disparity in China from 1983 to 1995. *Journal of Comparative Economics*. 27(4): 686-701.
18/ Kim, T. J. Knaap, G. J. 2001. The spatial dispersion of economic activities and development trends in China: 1952-1985. *The Annals of Regional Science*. 35(1): 39-57
19/ Keidel, A. J. 1995. *China: Regional Disparity*. Washington, DC: World Bank.
20/ Lardy, N. R. 1978. *Economic growth and income distribution in the People's Republic of China*. New York: Cambridge University Press.
21/ ---------------. 1980. Regional growth and income distribution in China, In R. F. Denberger (Ed.), *China's Development Experience in Comparative Perspective* (153-190). Cambridge: Harvard University Press.
22/ Li, X. Qiao, J. 2001. County level economic disparities of China in the 1990s. *Acta Geographica Sinica*. 56(2): 136-145.
23/ Lin, Y. Cai, F. et al. 1998. An analysis of regional gaps in China's economic transition. *Economic Research*

*Journal* (in Chinese). 6: 3-10.

24/ Lippit, V. D. 1987. *The Economic Development of China*. New York: ME Sharpe.

25/ Liu, Q. 2001. An analysis of convergence of China's economic growth. *Economic Research* (in Chinese). 6: 70-77.

26/ Long, G. N. & M. K. Ng 2001. The political economy of intra-provincial disparity in post-reform China: a case study of Jiansu province. *Geoforum*. 32: 215-234.

27/ Lu, D. Xue, F. et al. 1998. *The Report of China's Regional Development in 1997* (in Chinese). Beijing: The Commercial Press.

28/ Lyons, T. P. 1991. Inter-provincial disparities in China: output and consumption, 1952-1987. *Economic Development and Cultural Change*, 39(3): 471-506.

29/ ---------------. 1998. Intra-provincial disparity in China: Fujian Province, 1978-1995. *Economic Geography*, 74(3): 201-227.

30/ Mandelbrot, B. B. Wallis, J. R. 1968. Noah, Joseph and operational hydrology. *Water Resources Research* 4: 909-918.

31/ Mandelbrot, B. B. 1973. Formes nouvelles du hazard dans les science. *Economie Appliq* 26: 307-319.

32/ Masahisa F. Hu, D. 2001. Regional disparity in China 1985-1994: the effects of globalization and economic liberalization. *The Annals of Regional Science*. 35(1): 3-37.

33/ Roll, C. R., Jr. and Yeh, K. 1975. Balance in Coastal and Inland Industrial Development, in China. A Reassessment of the Economy, U.S.Congress Joint Economic Committee. Washington, D. C.: Government Printing Office.

34/ Rozelle, S. 1994. Rural industrialization and increasing inequality: emerging Patterns in China's reforming economy. *Journal of Comparative Economics*, 19(3): 362-391.

35/ Selden, M. 1988. *The Political Economy of Contemporary China*. New York: Sharpe.

36/ Song, D. 1998. Regional difference in economic development since the reform. *Quantitative & Technical Economics* (in Chinese), 3: 15-18.

37/ Theil, H. 1967. *Economics and Information Theory*. Amsterdam: North Holland.

38/ Tsui, K. Y. 1991. China's regional inequality, 1952-1985. *Journal of Comparative Economics*, 15(1): 1-21.

39/ -------------1998. Factor decomposition of Chinese rural income inequality: new methodology, empirical findings, and policy implications. *Journal of Comparative Economics*, 26(3): 502-528.

40/ Wei, H. 1992. The evolution pattern of regional income disparities in China. *Economic Research* (in Chinese), 4: 51-55.

41/ ----------1996. An Analysis of China's regional income disparities. *Economic Research* (in Chinese). 11,66-73.

42/ ----------1998. Theories in regional economic science. *Development Research* (in Chinese), 1, 34-38.

43/ Wei, H. Liu, K. 1994. The evolution of regional disparities in China: analysis and forecast. *Industrial Economics Research* (in Chinese), 4,28-36.

44/ Williamson, J. G. 1965. Regional inequality and the process of national development: A description of patterns. *Economic Development and Cultural Change* 13 (4): 3-84.

45/ Yang, D. 1990. Patterns of China's regional development strategy. *China Quarterly*. 122: 231-257.

46/ Yang, K. 1994. The evolution of the regional economic disparities in China. *Economic Research Journal* (in Chinese). 12: 28-33.

47/ Yang, W. 1992. *Empirical studies. Economic Research* (in Chinese). 1: 70-74.

48/ Yao, S. and Zhang, Z. 2001. On regional disparity and diverging clubs: a case study of contemporary China. *Journal of Comparative Economics*. 29(3): 466-484.

49/ Ying, L. 1999. China's changing regional disparity during the reform period. *Economic Geography*. 75(1): 59-70.

50/ Yuan, G. 1996. Regional economic disparity and macroeconomic fluctuation. *Economic Research* (in Chinese), 10: 49-56.